\documentclass[aps,preprint,nofootinbib]{revtex4}%
\usepackage{amsfonts}
\usepackage{amsmath}
\usepackage{amssymb}
\usepackage{float}
\usepackage{graphicx}%
\providecommand{\U}[1]{\protect\rule{.1in}{.1in}}

\begin{document}
\title{Homogeneous Black Strings in Einstein-Gauss-Bonnet with Horndeski hair and beyond}
\author{Adolfo Cisterna$^{1}$, Sebasti\'{a}n Fuenzalida$^{2}$, Marcela Lagos$^{2}$ and
Julio Oliva$^{2}$}
\affiliation{$^{1}$Centro de Ingenier\'{\i}a y Desarrollo Sustentable, Facultad de
Ingenier\'{\i}a, Universidad Central de Chile, Santa Isabel 1186, 8330601
Santiago, Chile}
\affiliation{$^{2}$Departamento de F\'{\i}sica, Universidad de Concepci\'{o}n, Casilla,
160-C, Concepci\'{o}n, Chile.}
\email{adolfo.cisterna@ucentral.cl, sfuenzalida@udec.cl, marcelagos@udec.cl, julioolivazapata@gmail.cl}

\begin{abstract}
In this paper we construct new exact solutions in Einstein-Gauss-Bonnet and
Lovelock gravity, describing asymptotically flat black strings. The solutions
exist also under the inclusion of a cosmological term in the action, and are
supported by scalar fields with finite energy density, which are linear along
the extended direction and have kinetic terms constructed out from Lovelock
tensors. The divergenceless nature of the Lovelock tensors in the kinetic
terms ensures that the whole theory is second order. For spherically,
hyperbolic and planar symmetric spacetimes on the string, we obtain an
effective Wheeler's polynomial which determines the lapse function up to an
algebraic equation. For the sake of concreteness, we explicitly show the
existence of a family of asymptotically flat black strings in six dimensions, as well as asymptotically AdS$_{5}\times%
\mathbb{R}
$ black string solutions and compute the temperature, mass density and entropy
density. We compute the latter by Wald's formula and show that it receives a
contribution from the non-minimal kinetic coupling of the matter part,
shifting the one-quarter factor coming from the Einstein term, on top of the
usual non areal contribution arising from the quadratic Gauss-Bonnet term.
Finally, for a special value of the couplings of the theory in six dimensions,
we construct strings that contain asymptotically AdS wormholes as well as
rotating solutions on the transverse section. By including more scalars the
strings can be extended to $p$-branes, in arbitrary dimensions.

\end{abstract}
\maketitle

\section{Introduction}

Higher dimensional General Relativity (GR) possesses a broader spectrum of
black hole solutions as compared with its four dimensional formulation
\cite{Obers,Horo}. Indeed, four dimensional GR is constrained by uniqueness
theorems \cite{UT1,UT2,UT3} that ensure that any black hole solution of the
theory is contained in the Kerr family \cite{Kerr:1963ud,Debney:1969zz}.
Moreover, topological restrictions allow only for horizons with spherical
topology \cite{Fried}. The spectrum of solutions is limited not only
quantitatively but also qualitatively, setting the final state of black hole
collapse to be described only by a small set of parameters
\cite{Ruffini:1971bza}. On the other hand it is well-known that gravity in
higher dimensions admits spacetimes with horizons that can have more general
topologies than that of the $\left(  d-2\right)  $-sphere \cite{Obers,Horo},
being the existence of black strings, a black hole solution with horizon
structure $S^{(d-3)}\times R$, the simplest counterexample \cite{UT1,UT2,UT3}
as it coexists with the Schwarzschild-Tangherlini black hole \cite{Tangher}.
Black strings also paved the road for the construction of more sophisticated
asymptotically flat solutions with non-spherical topology such as black rings
\cite{Empa} and diverse black object solutions \cite{Horo}, demonstrating how
topological restrictions \cite{Fried} are weakened in higher dimensions. It
was shown that black strings are affected by Gregory-Laflamme (GL) instability
\cite{GL1,GL2}, a long-wavelength perturbative instability triggered by a mode
that travels along the extended direction of the horizon and, moreover in
dimension five the numerical simulations indicate that the instability ends in
the formation of naked singularities \cite{PLuis1,PLuis2}, representing an
explicit failure of the cosmic censorship in higher dimensions \cite{RMW}
\footnote{This feature is also present in the black ring \cite{Dias:2009iu}
and the Myers-Perry solution in six dimensions \cite{Dias:2010eu}.}.\newline
Black strings in GR in vacuum are easy to construct. In fact they are obtained
by a cylindrical oxidation of the Schwarzschild black hole in $d$ dimensions
by the inclusion of $p$ extra flat coordinates. This can be realized by seeing
that the equations of motion along the extended directions are compatible with
the field equations on the $p$-brane, due to the fact that the involved
curvature quantities vanish on these coordinates. \newline Notwithstanding it
is not hard to find simple scenarios in which the construction of analytic
black strings fails. Indeed, the mere inclusion of a cosmological constant
spoils the existence of cylindrically extended black strings, since the
compatibility of the equations of motion on the flat coordinates with the
trace of the field equations along the brane forces $\Lambda$ to vanish. This
implies that there is no simple oxidation of the Schwarzschild (A)dS black
hole\footnote{Numerical as well as perturbative solutions have been
constructed in \cite{numads} in GR as well as in gauged supergravity in five
dimensions.}. \newline In \cite{Cisterna:2017qrb} two of the authors have
shown that GR with a negative cosmological constant does admit exact,
homogeneous black string solutions if each extended flat coordinate is dressed
with a massless, minimally coupled scalar field that depends exclusively on
that coordinate\footnote{Previously, AdS black strings in GR were constructed
considering warped spacetimes \cite{Chamblin:1999by} providing non-homogenous
configurations.}. To see this explicitly let us take
the following black string ansatz in which a $d$-dimensional black hole is
oxidated by including $p$ flat directions
\begin{equation}
ds^{2}=-F(r)dt^{2}+\frac{dr^{2}}{F(r)}+r^{2}d\Sigma_{d-2,K}^{2}%
+\delta_{ij}dx^{i}dx^{j}\ ,\label{ansatzbs}%
\end{equation}
where $i,j=1,2,\ldots,p$ and $d\Sigma_{d-2,K}^{2}$ stands for a
$(d-2)$-dimensional Euclidean manifold of constant curvature $K=0,\pm1$.
By virtue of the ansatz (\ref{ansatzbs}), the massless Klein-Gordon equations
deliver scalar fields that depend linearly on the extended flat directions
\begin{equation}
\psi_{(j)}(x^{j})=\lambda_{(j)}x^{j}\ ,\quad\text{no sum over $j$%
.}\label{axion}%
\end{equation}
the $\lambda_{(j)}$ being integration constants. These scalars cure the
incompatibility between the equations of motion on the brane and the extended
directions leading to the metric
\begin{equation}
F(r)=K-\frac{2\mu}{r^{d-3}}-\frac{2\Lambda r^{2}}{\left(  d-1\right)
\left(  d+p-2\right)  }\ ,\label{MU}%
\end{equation}
where compatibility is ensured provided
\begin{equation}
\lambda_{(j)}^{2}=\lambda^{2}=-\frac{4\Lambda}{\left(  d+p-2\right)  }\ .
\end{equation}
This solution represents the black string version of the Schwarzschild-AdS
black hole and shows that the three-dimensional BTZ black hole can be also
uplifted to a higher dimensional black string\footnote{This can also be
achieved in the Einstein-Skyrme system \cite{Astorino:2018dtr} and even more
by using this approach, diverse BTZ black strings have been recently
constructed in theories with non-trivial torsion \cite{Cisterna:2018jsx}.}.
\newline A natural question arises: As higher dimensional objects, do
homogeneous and cylindrically extended black strings and black $p$-branes
exist in higher curvature extensions of GR? \newline In this paper we answer
this question affirmatively for the Einstein-Gauss-Bonnet as well as for
general Lovelock theories with arbitrary values of the coupling constants. To
successfully apply the procedure before described, we shall observe that the
extended directions must be dressed with scalar fields of the type
(\ref{axion}) that have non-minimal kinetic couplings, such that the extra
higher curvature terms involved in Lovelock theories do not break the
compatibility of the field equations. Such couplings must fulfil the following
requirements: They must enjoy shift symmetry, which allows the inclusion of
non-minimal kinetic couplings. Secondly, their contribution to the equations
of motion must be of the same type as that of the Lovelock terms under
consideration. These types of non-minimal kinetic couplings arise naturally in
Galileon/Horndeski theory namely, the most general scalar-tensor theory with
second order field equations for both, the metric and the scalar field.
Originally constructed in the early seventies, Horndeski theory
\cite{Horndeski} has re-emerged after the appearances of Galileon theories
\cite{Gal1,Gal2,Gal3} mostly motivated by their several applications in
cosmology \cite{cosmo01,cosmo02,cosmo03}. The Horndeski Lagrangian, which has
been constructed in four dimensions, has a non-minimal kinetic sector given
by
\begin{equation}
\mathcal{L}\sim G_{AB}\partial^{A}\chi\partial^{B}\chi\ ,\label{ekc}%
\end{equation}
where $G_{AB}$ is the Einstein tensor. This term was extensively considered
not only in cosmology \cite{cosmo1,cosmo2,cosmo3,cosmo4,cosmo5} but also in
the study of compact objects such as black holes and neutron stars
\cite{bh1,bh2,bh3,bh4,bh5,bh6,bh7,Cisterna:2015yla,Cisterna:2016vdx}. We will
see that the higher dimensional generalizations of this term are precisely the
kind of couplings we need to consider in order to construct homogeneous black
strings in Lovelock gravity. The shift invariant scalars will provide the
dress we need in order to ensure compatibility of the equations of
motion.\newline This paper is organized as follows: To fix ideas, Section II
is destined to construct homogeneous black strings in Einstein-Gauss-Bonnet
theory dressed by two scalar fields, a minimally coupled one accounting for
the inclusion of the cosmological constant and a non-minimally coupled one
accounting for the new $R^{2}$ curvature terms we are including. We present the equations in arbitrary dimension $D=d+1$, nevertheless for the sake
of concreteness we focus on the minimal dimension allowing for new black strings which in this case is $D=(1+4)+1$ and analyze the
causal structures as well as the thermal properties of the black strings.
Section III is devoted to extend the previous results to theories containing
arbitrarily higher curvature terms in the Lovelock family. Remarkably, we show
that there is a pattern that allows to construct homogeneous AdS black strings
in Lovelock theories supported by a family of scalars with generalized
non-minimal kinetic terms, leading to a Wheeler's like polynomial with
effective couplings. Assuming the existence of an event horizon, we compute
the temperature, as well as the entropy of the black hole. In Section IV we
revisit the Einstein-Gauss-Bonnet theory and show that for special values of
the matter couplings, if the transverse section of the black strings and
$p$-branes is five dimensional, the theory allows for wormholes as well as
rotating solutions. The latter leads to the first example of a rotating
solution in Einstein-Gauss-Bonnet theory in arbitrary dimensions. Finally in
Section V we outline our conclusions and further developments that can follow
this work.

\section{Homogeneous black strings in Einstein-Gauss-Bonnet theory}

Let's consider the Einstein-Gauss-Bonnet theory in arbitrary dimension coupled
to two scalar fields as follows:%

\begin{equation}
I=\int d^{D}x\sqrt{-g}\left(  R-2\Lambda_{0}+\alpha\mathcal{L}_{GB}-\frac
{1}{2}g_{AB}\partial^{A}\psi\partial^{B}\psi+\frac{\gamma}{2}G_{AB}%
\partial^{A}\chi\partial^{B}\chi\right)  \ ,\label{actionegb}%
\end{equation}
\newline where $\mathcal{L}_{GB}:=R^{2}-4R_{AB}R^{AB}+R_{ABCD}R^{ABCD}$
defines the Gauss-Bonnet term. Performing the variations respect to the metric
and the scalar fields we obtain the following set of field equations
\begin{align}
G_{AB}+\Lambda_{0}g_{AB}+\alpha H_{AB\ } &  =T_{AB}^{\left(  1\right)
}+T_{AB}^{\left(  2\right)  }\ ,\\
g^{AB}\nabla_{A}\nabla_{B}\psi &  =0\ ,\\
\gamma\ G^{AB}\nabla_{A}\nabla_{B}\chi &  =0\ ,\label{eccampo}%
\end{align}
where the Gauss-Bonnet tensor is given by%
\[
H_{AB}=2RR_{AB}-4R_{ACBD}R^{CD}+2R_{ACDE}R_{B}^{\ CDE}-4R_{AC}R_{B}%
^{\ \ C}-\frac{1}{2}g_{AB}\mathcal{L}_{GB}\ ,
\]
and the energy-momentum tensors for the scalars read%
\begin{align*}
T_{AB}^{\left(  1\right)  }   =&\frac{1}{2}\left(  \partial_{A}\psi
\partial_{B}\psi-\frac{1}{2}g_{AB}\left(  \partial\psi\right)  ^{2}\right)
\ ,\\
T_{AB}^{\left(  2\right)  }   =&\frac{\gamma}{2}\left(  \frac{1}{2}%
\partial_{A}\chi\partial_{B}\chi R-2\partial_{C}\chi\partial_{(A}\chi
R_{B)}^{\ \ C}-\partial_{C}\chi\partial_{D}\chi R_{A\ \ B}^{\ C\ \ \ D}%
-\nabla_{A}\nabla^{C}\chi\nabla_{B}\nabla_{C}\chi+\nabla_{A}\nabla_{B}%
\chi\square\chi\right.  \\
&  \left.  +\frac{1}{2}G_{AB}\left(  \partial\chi\right)  ^{2}-g_{AB}\left[
-\frac{1}{2}\nabla^{C}\nabla^{D}\chi\nabla_{C}\nabla_{D}\chi+\frac{1}%
{2}\left(  \square\chi\right)  ^{2}-\partial_{C}\chi\partial_{D}\chi
R^{CD}\right]  \right)  \ .
\end{align*}
Note that the equations (\ref{eccampo}) are linear on the scalars.

On a metric of the form%
\begin{equation}
ds^{2}=d\tilde{s}_{d}^{2}+dz^{2}\ ,\label{bsfullD}%
\end{equation}
and scalars that depend only on the extended direction $z$, the equations for
the scalars imply a linear dependence on $z$. The shift symmetry of the scalar
field theories can be used to set to zero the additive integration constants
that appears in both scalars. So we have%
\begin{equation}
\psi\left(  z\right)  =c_{0}z\text{ and }\chi\left(  z\right)  =c_{1}z\ ,
\end{equation}
with $c_{0}$ and $c_{1},$ integration constants. Then, the trace of the field
equations on the $d$-dimensional manifold and the equation along the $z$
direction respectively read%
\begin{align}
\mathcal{E}_{1}  & :=\left(  \Lambda_{0}+\frac{c_{0}^{2}}{4}\right)  d+\left(
1-\frac{\gamma}{4}c_{1}^{2}\right)  \left(  1-\frac{d}{2}\right)
\tilde{R}+\alpha\left(  2-\frac{d}{2}\right)  \mathcal{\tilde{L}}_{GB}=0\ ,\\
\mathcal{E}_{2}  & :=\left(  \Lambda_{0}-\frac{c_{0}^{2}}{4}\right)  -\frac
{1}{2}\left(  1+\frac{\gamma}{4}c_{1}^{2}\right) \tilde{R}-\frac{\alpha}%
{2}\mathcal{\tilde{L}}_{GB}=0\ ,
\end{align}
where $\tilde{R}$ and $\mathcal{\tilde{L}}_{GB}$ are the intrinsic Ricci scalar and Gauss-Bonnet combination of the $d$-dimensional manifold with line element $d\tilde{s}_{d}$. To avoid incompatibilities these two equations must be proportional term by
term, i.e. $\mathcal{E}_{1}=\xi\mathcal{E}_{2}$ and therefore, when the higher
curvature Gauss-Bonnet term is present, one obtains that the proportionality
constant must be fixed as well as the integration constants of the scalars,
leading to\footnote{Note that when $\alpha=0$, one obtains only two equations
for the three constant $\xi$, $c_{0}$ and $c_{1}$, leading to $c_{0}%
^{2}=-\frac{4\Lambda_{0}\left(  \xi d-1\right)  }{\xi d+1}$ and $c_{1}%
^{2}=\frac{4}{\gamma}\frac{\left(  \left(  d-2\right)  \xi-1\right)  }{\left(
\left(  d-2\right)  \xi+1\right)  }$ for an arbitrary $\xi$. Since we are
interested in the inclusion of higher derivative terms, we do not elaborated
further in the case $\alpha=0$.}%
\begin{equation}
\xi=\frac{1}{d-4},\ c_{0}^{2}=-\frac{8\Lambda_{0}}{d-2}\text{ and }c_{1}%
^{2}=\frac{4}{\left(  d-3\right)  \gamma}\ .
\end{equation}

Under these conditions, we will have a solution of the theory (\ref{actionegb}) in
arbitrary dimension $D=d+1$ for the scalars%
\begin{equation}
\psi\left(  z\right)  =\sqrt{-\frac{8\Lambda_{0}}{d-2}}z\text{ and }%
\chi\left(  z\right)  =\sqrt{\frac{4}{\left(  d-3\right)  \gamma}}z\ ,
\end{equation}
provided the $d-$dimensional metric $d\tilde{s}_{d}^{2}$ fulfils the
Einstein-Gauss-Bonnet field equations with the following rescaled couplings%
\begin{equation}
\frac{\left(  d-4\right)  \left(  d-3\right)  \Lambda_{0}}{\left(  d-2\right)
}\tilde{g}_{\mu\nu}+\left(  d-4\right)  \tilde{G}_{\mu\nu}+\left(  d-3\right)  \alpha
\tilde{H}_{\mu\nu}=0\ .\label{reducedEGB}%
\end{equation}
Here $\tilde{g}_{\mu\nu}$, $\tilde{G}_{\mu\nu}$ and $\tilde{H}_{\mu\nu}$ are, respectively, the intrinsic metric,
Einstein tensor and Gauss-Bonnet tensor for the $d-$dimensional metric
$d\tilde{s}_{d}^{2}$ on the constant $z$ section of the black string\footnote{Note that the equations that determine the geometry (\ref{reducedEGB}) do not depend on $\gamma$, since the energy-momentum tensor is proportional to the coupling $\gamma$ and at the same time is quadratic in the field $\chi\sim\gamma^{-1/2}$.} (\ref{bsfullD}).
Since $\tilde{H}_{\mu\nu}$ vanishes identically in dimensions $d\leq4$, to have a
non-vanishing contribution from the Gauss-Bonnet tensor, we need $d\geq5$ (in
$d=4$ the equation (\ref{reducedEGB}) is identically fulfilled for any metric
and the system is degenerate at such point).

For simplicity, let's focus on the six dimensional case and consider the
homogeneous black string metric%
\begin{equation}
ds_{6}^{2}=-f\left(  r\right)  dt^{2}+\frac{dr^{2}}{f\left(  r\right)  }%
+r^{2}d\Sigma_{3,K}^{2}+dz^{2}\ ,
\end{equation}
where $d\Sigma_{3,K}$ is the line element of an Euclidean, three-dimensional
constant curvature manifold of curvature $K$, and assume that the scalars
depend only on the $z$ direction.

The explicit $z$-dependence of the scalars is fixed by the scalar equations to
be linear, giving rise to%
\begin{equation}
\psi\left(  z\right)  =\sqrt{-\frac{8\Lambda_{0}}{3}}z\text{ \ \ and \ \ }\chi\left(
z\right)  =\sqrt{\frac{2}{\gamma}}z\ ,\label{escalares}%
\end{equation}
and the metric function has to solve the following quadratic, polynomial
equation%
\begin{equation}
36\alpha\left(K-f\left(r\right)\right)^2+r^2\left(9K-\Lambda_0 r^2-9f\left(r\right)\right)=\frac{6m}{V^{\left(K\right)}_{3}}
\end{equation}
where $m$ is an integration constant and $V^{\left( K\right) }_{3}$ stands for the volume of manifold with line element $d\Sigma_{K,3}$.

For vanishing cosmological term in the action ($\Lambda_{0}=0$), the minimally
coupled scalar vanishes and as a solution of the polynomial equation one
obtain%
\begin{equation}
f\left(  r\right)  =K+\frac{r^{2}}{8\alpha}\left(  1\pm\sqrt{1+\frac
{32}{3V^{\left(K\right)}_{3}}\frac{m\alpha}{r^{4}}}\right)  \ .
\end{equation}
Consequently, for the spherically symmetric case, the metric describes an
asymptotically flat, homogeneous black string, with a regular horizon, in the six-dimensional
Einstein-Gauss-Bonnet theory supported by a scalar field with an
Einstein-kinetic term. These are the first known asymptotically flat, analytic
and homogeneous black string solutions in general relativity with a
Gauss-Bonnet term.

\bigskip

Considering a non-vanishing bare cosmological term, $\Lambda_{0}\neq 0$, the solutions of the polynomial
reads%
\begin{equation}
f\left(  r\right)  =K+\frac{r^{2}}{8\alpha}\left(  1\pm\sqrt{1+\frac{16}{9}\alpha\Lambda_0+\frac{32}{3V^{\left(K\right)}_{3}}\frac{m\alpha}{r^{4}}}\right)  \ .
\label{f7}%
\end{equation}
As expected, this has the form of the topological extension of the
Boulware-Deser metric with a cosmological constant \cite{BD}.

From the expression (\ref{escalares}) we see that the bare cosmological
constant must fulfil $\Lambda_{0}\leq0$, therefore, for simplicity and without
loosing generality, let us set%
\begin{equation}
\Lambda_{0}=-1\ .
\end{equation}

For positive values of the Gauss-Bonnet coupling ($\alpha>0$) and spherically
symmetric spacetimes on the brane ($K=1$) only the branch with the negative
sign in (\ref{f7}) might lead to an event horizon. In this case, the
asymptotic behavior of the metric is%
\begin{align}
f\left(  r\right)   &  =1+\frac{r^{2}}{8\alpha}\left(  1-\sqrt{1-\frac
{16}{9}\alpha}\right)  +\mathcal{O}\left(  Mr^{-2}\right)  \\
&  =1+\frac{r^{2}}{l^{2}}+\mathcal{O}\left(  Mr^{-2}\right)  \ ,
\end{align}
which induces an upper bound on the Gauss-Bonnet coupling $0<\alpha<\frac
{9}{16}$, where we have defined%
\begin{equation}
\frac{1}{l^{2}}=\frac{1}{8\alpha}\left(  1-\sqrt{1-\frac
{16}{9}\alpha}\right)  \ .
\end{equation}
Then, the asymptotic behavior of the metric on the brane is that of the
$AdS_{5}$ spacetime and the subleading term is controlled by the integration
constant $m$, leading to a finite mass contribution (mass density if one
considers the extended direction). Considering $\alpha$ positive and $l^{2}$
to be the curvature radius of the brane at infinity, leads to the restriction
$\frac{9}{2}<l^{2}<9$.

Wald's formula \cite{Wald:1993nt} can be used to obtain the following entropy
per unit length of the string%
\begin{equation}
s=4\pi^{3}r_{+}^{3}+96\pi^{3}r_{+}\alpha\ ,
\end{equation}
where $r_{+}$ is the horizon radius. Here we note that the Einstein
contribution to the entropy receives a correction from the matter sector.
Since in this case the Bekenstein-Hawking entropy should be $\frac{A}%
{4G}=8\pi^{3}r_{+}^{3}L$ (with $16\pi G=1$). As expected on
dimensional grounds, the term proportional to the Gauss-Bonnet coupling
depends linearly on the horizon radius. As occurs for black holes, the
entropy of the planar horizons do not receive corrections from the higher
curvature terms.

The temperature is fixed in order to obtain a smooth Euclidean continuation on
the horizon and is given by%
\begin{equation}
T=\frac{2r^{3}_{+}+9r_{+}}{2\pi\left(9r^2_{+}-2l^4+18l^2\right)  }\ .
\end{equation}
One can see that, identifying the integration constant $m$ with the mass density, the
thermodynamical quantities fulfil the first law%
\begin{equation}
dm=Tds\ .
\end{equation}
Since the integration constants of the scalars are fixed in terms of the
couplings, it is natural to expect that the first law does not receive
contributions from the matter sector. A plot with different profiles for the
function $f\left(  r\right)  $ is presented in figure 1, for both the
asymptotically flat (left panel) and asymptotically AdS solutions (right panel).

\begin{figure}[ptb]
\centering
\includegraphics[width=0.45\textwidth]{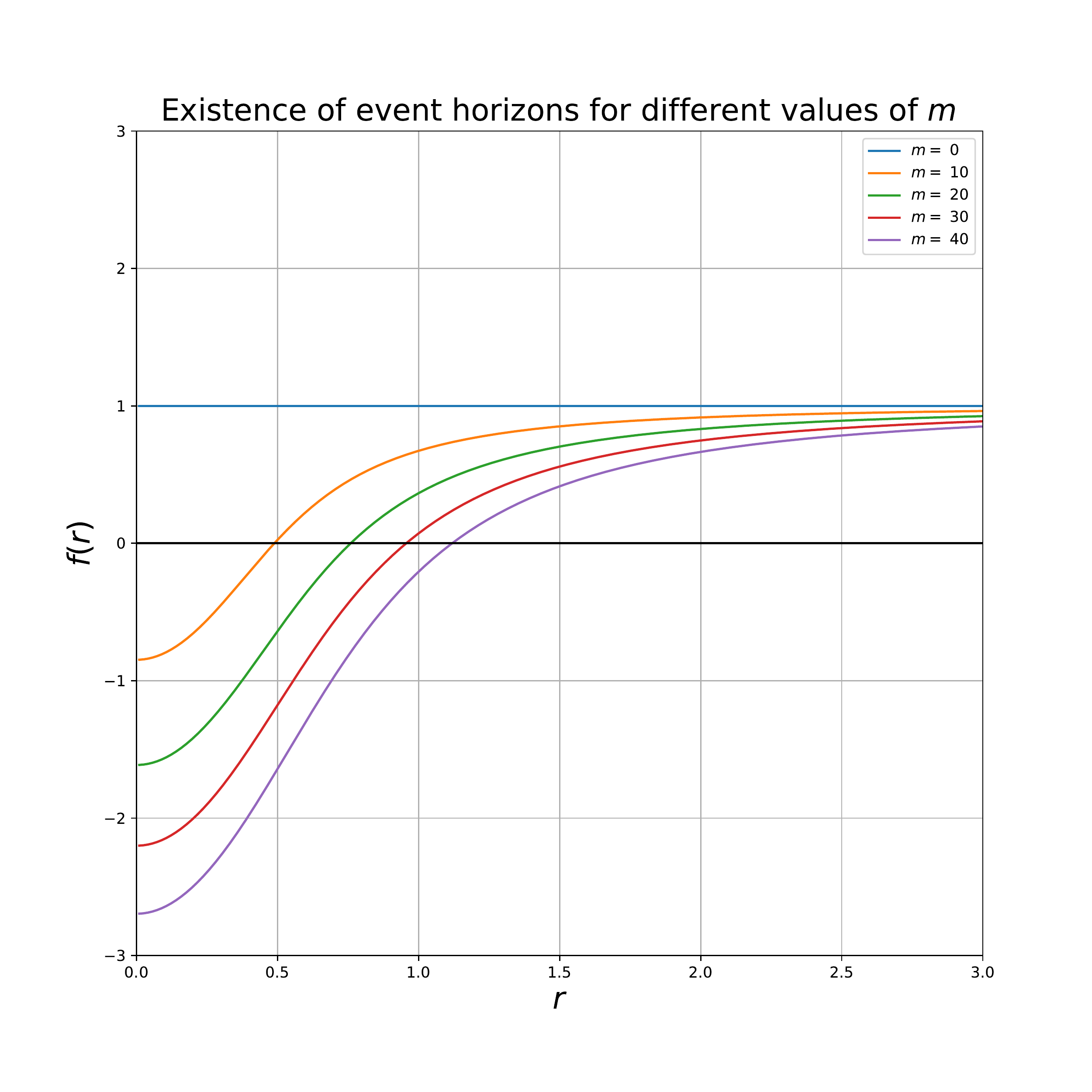}\qquad
\includegraphics[width=0.45\textwidth]{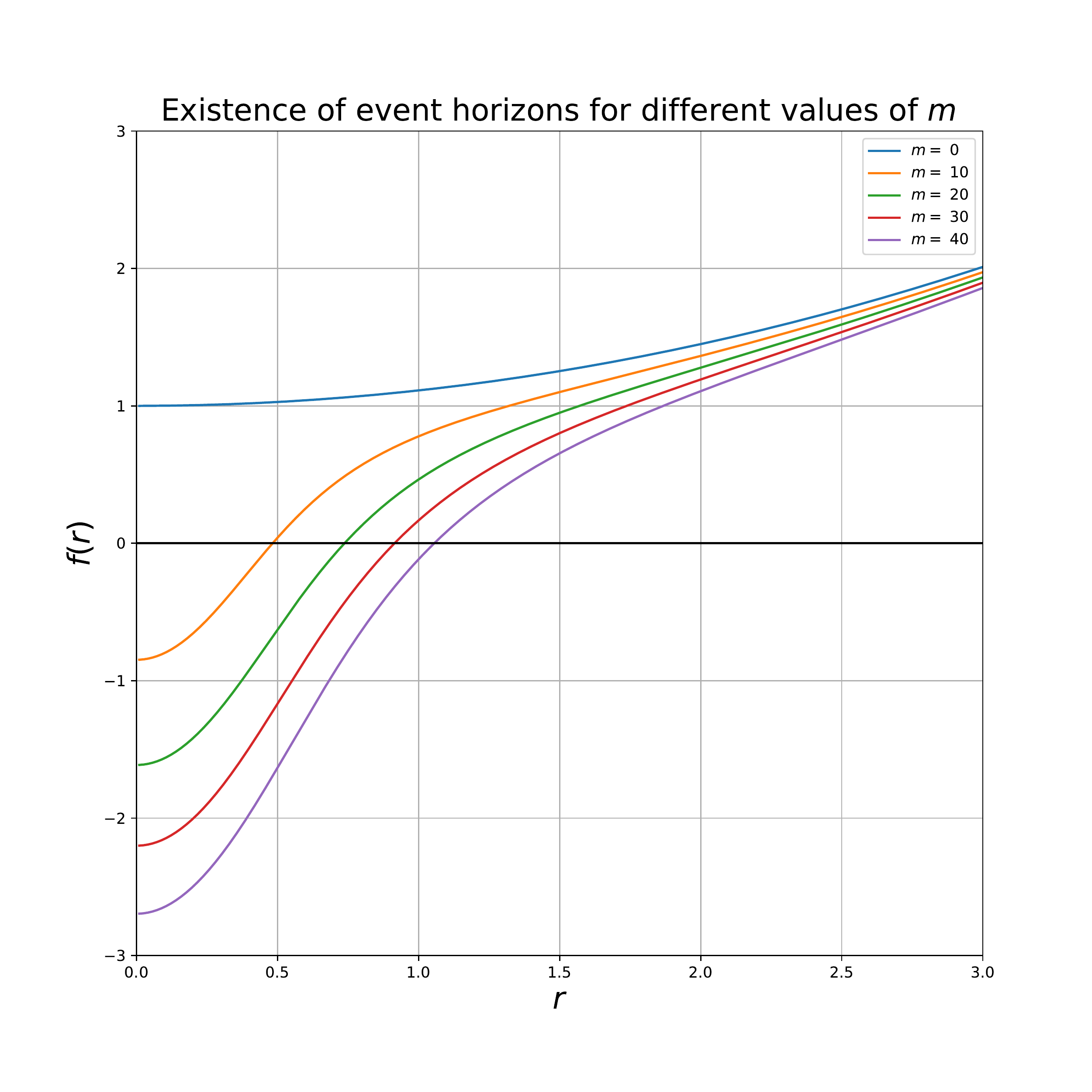} \caption{Here we plot the metric
function $f(r)$ coming from the effective Wheeler's polynomial, for different
values of the integration constant $m$. The black holes in the left panel lead
to asymptotically flat black strings in Einstein-Gauss-Bonnet theory in six dimensions, while the right panel lead to asymptotically $AdS_{5}\times R$
black strings. Note that the metric function $f(r)$ is regular at the origin, as it occurs for EGB in five dimensions in vacuum.}%
\label{fig1}%
\end{figure}\bigskip

We have therefore shown that the inclusion of the non-minimal Einstein-kinetic
coupling (\ref{ekc}) allows to construct black strings in
Einstein-Gauss-Bonnet theory, for arbitrary values of the coupling constants
of the theory.

\bigskip

As mentioned above, the obstruction to the existence of cylindrically oxidated
solutions comes from the incompatibility between the trace of the field
equations on the brane and the equation along the extended direction. The
scalar fields $\psi$ and $\chi$ provide a natural manner to circumvent this
incompatibility even in the asymptotically flat case. Below, we give a
detailed explanation of the mechanism behind the existence of these solutions
for general Lovelock theories, and show that the results can be extended
beyond Einstein-Gauss-Bonnet by the inclusion of scalars with non-minimal
kinetic coupling to Lovelock tensors, that naturally extend (\ref{ekc}).

\section{homogeneous black strings in general Lovelock theory}

Let us supplement the Lovelock action by quadratic scalars in the following form%

\begin{equation}
I[g_{AB},\phi_{\left(  k\right)  }]=\int d^{D}x\left[  \sum_{k=0}^{n+1}%
\alpha_{k}\mathcal{L}^{(k)}+\sum_{k=0}^{n}\beta_{k}\mathcal{L}_{\text{matter}%
}^{(k)}\right]  \ , \label{resultados-principales:accion-llh}%
\end{equation}
where the Lovelock Lagrangians are given by%

\begin{equation}
\mathcal{L}^{\left(  k\right)  }=\frac{\sqrt{-g}}{2^{k}}\delta_{B_{1}\cdots
B_{2k}}^{A_{1}\cdots A_{2k}}R^{B_{1}B_{2}}{}{}_{A_{1}A_{2}}\cdots
R^{B_{2k-1}B_{2k}}{}{}_{A_{2k-1}A_{2k}}\ ,
\label{resultados-principales:lagrangiano-ll}%
\end{equation}
and each of the matter fields $\phi_{\left(  k\right)  }=\left\{  \psi
,\chi,...\right\}  $ with $k=0,1,...,n\ $, have different dynamics controlled
by the Lagrangian%

\begin{align}
\mathcal{L}_{\text{matter}}^{\left(  k\right)  } &  =\frac{\sqrt{-g}}{2^{k}%
}E_{CD}^{(k)}\nabla^{C}\phi_{(k)}\nabla^{D}\phi_{(k)}\nonumber\\
&  =-\frac{\sqrt{-g}}{2^{2k+1}}\nabla_{C}\phi_{(k)}\nabla^{D}\phi_{(k)}%
\delta_{DB_{1}\cdots B_{2k}}^{CA_{1}\cdots A_{2k}}R^{B_{1}B_{2}}{}{}%
_{A_{1}A_{2}}\cdots R^{B_{2k-1}B_{2k}}{}{}_{A_{2k-1}A_{2k}}\ ,
\end{align}
which couples the $k$-th scalar directly to the $k$-th order Lovelock tensor
$E_{CD}^{(k)}$ (\ref{resultados-principales:tensor-lovelock}). The Lovelock
tensor has a divergence that vanishes identically since it corresponds to the
Euler-Lagrange derivative with respect to the metric of a diffeomorphism
invariant action. The $\alpha_{k}$ are the dimensionful Lovelock couplings and
$\beta_{k}$ are the matter couplings.

Varying (\ref{resultados-principales:accion-llh}) with respect metric $g^{AB}$
we obtain the field equations%

\begin{equation}
\mathcal{E}_{AB}:=\sum_{k=0}^{n+1}\alpha_{k}E_{AB}^{(k)}-\sum_{k=0}^{n}%
\frac{\beta_{k}}{2^{2k+1}}T_{AB}^{(k)}=0\ ,
\label{resultados-principales:ec-lh}%
\end{equation}
with the Lovelock tensor of order $k$ defined as%

\begin{equation}
E_{AB}^{(k)}:=-\frac{1}{2^{k+1}}g_{\left(  A\right\vert C}\delta_{\left\vert
B\right)  B_{1}\cdots B_{2k}}^{CA_{1}\cdots A_{2k}}R^{B_{1}B_{2}}{}{}%
_{A_{1}A_{2}}\cdots R^{B_{2k-1}B_{2k}}{}{}_{A_{2k-1}A_{2k}}\ .
\label{resultados-principales:tensor-lovelock}%
\end{equation}

Furthermore, the energy-momentum tensor associated with the $k$-th scalar
$\phi_{\left(  k\right)  }$ is given by%
\begin{align}
T_{AB}^{(k)}=  &  -\frac{1}{2}g_{AB}\nabla_{C}\phi_{(k)}\nabla^{D}\phi
_{(k)}\delta_{DB_{1}\cdots B_{2k}}^{CA_{1}\cdots A_{2k}}R^{B_{1}B_{2}}{}%
{}_{A_{1}A_{2}}\cdots R^{B_{2k-1}B_{2k}}{}{}_{A_{2k-1}A_{2k}}\nonumber\\
&  +\nabla_{C}\phi_{(k)}\nabla_{\left(  A\right\vert }\phi_{(k)}%
\delta_{\left\vert B\right)  B_{1}\cdots B_{2k}}^{CA_{1}\cdots A_{2k}}%
R^{B_{1}B_{2}}{}{}_{A_{1}A_{2}}\cdots R^{B_{2k-1}B_{2k}}{}{}_{A_{2k-1}A_{2k}%
}\nonumber\\
&  +k\nabla_{C}\phi_{(k)}\nabla^{D}\phi_{(k)}\delta_{DB_{1}\cdots\left(
A\right\vert }^{CA_{1}\cdots A_{2k}}R^{B_{1}B_{2}}{}{}_{A_{1}A_{2}}\cdots
R^{B_{2k-3}B_{2k-2}}{}{}_{A_{2k-3}A_{2k-2}}R^{B_{2k-1}}{}_{\left\vert
B\right)  A_{2k-1}A_{2k}}\nonumber\\
&  +2k\nabla^{B_{2k-1}}\nabla_{C}\phi_{(k)}\nabla_{A_{2k-1}}\nabla^{D}%
\phi_{(k)}\delta_{DB_{1}\cdots\left(  B\right\vert }^{CA_{1}\cdots A_{2k}%
}R^{B_{1}B_{2}}{}{}_{A_{1}A_{2}}\cdots R^{B_{2k-3}B_{2k-2}}{}{}_{A_{2k-3}%
A_{2k-2}}g_{\left\vert A\right)  A_{2k}}\nonumber\\
&  +k\nabla_{C}\phi_{(k)}\nabla^{E}\phi_{(k)}\delta_{DB_{1}\cdots\left(
B\right\vert }^{CA_{1}\cdots A_{2k}}R^{B_{1}B_{2}}{}{}_{A_{1}A_{2}}\cdots
R^{B_{2k-3}B_{2k-2}}{}{}_{A_{2k-3}A_{2k-2}}R_{A_{2k-1}E}{}{}^{B_{2k-1}%
D}g_{\left\vert A\right)  A_{2k}}\ .
\end{align}

On the other hand, by varying the action with respect to field $\phi_{\left(
k\right)  }$, we obtain the field equations%

\begin{equation}
\beta_{k}E_{AB}^{\left(  k\right)  }\nabla^{A}\nabla^{B}\phi_{(k)}=0\text{
with }%
k=0,...,n\ .\label{resultados-principales:ecuaciones-movimiento-generales-campo-phi}%
\end{equation}
Below we will show that these equations admit homogeneous black strings by
requiring the trace of the equations on the brane to be compatible with the
equation along the extended direction.

We consider the ansatz%

\begin{equation}
ds^{2}=d\tilde{s}_{d}^{2}+dz^{2}\ ,
\label{resultados-principales:ansatz-metrica}%
\end{equation}
where%

\begin{equation}
d\tilde{s}_{d}^{2}=-f(r)dt^{2}+\frac{dr^{2}}%
{f(r)}+r^{2}d\Sigma_{K,d-2}^{2}\ ,
\end{equation}
and the scalar fields depending only on the extended coordinate. The equations
for the fields
(\ref{resultados-principales:ecuaciones-movimiento-generales-campo-phi}) imply
that the scalars are linear in the extended direction, and the invariance
under constant shifts of the matter actions can be used to write%
\begin{equation}
\phi_{(k)}=c_{\left(  k\right)  }%
z\ .\label{resultados-principales:ansatz-campo-escalar}%
\end{equation}
Here $d\Sigma_{K,d-2}$ denotes the line element of an Euclidean
$\left(  d-2\right)  -$dimensional manifold with constant curvature
$K=\pm1,0\ .$

We will split the $D$-dimensional indices $\left\{  A,B,C,...\right\}  $ in
$\left\{  \mu,\nu,...\right\}  $ that run on the brane with line element
$d\tilde{s}_{d}$ and $z$ along the extended direction. Quantities
intrinsically defined on the $d$-dimensional manifold will have tildes on top.

To explain the previously mentioned compatibility, we will be as explicit as
possible in what follows. The field equations
(\ref{resultados-principales:ec-lh}) read%

\begin{align}
\alpha_{0}  &  E^{(0)}{}^{A}{}_{B}+\alpha_{1}E^{(1)}{}^{A}{}_{B}+\alpha
_{2}E^{(2)}{}^{A}{}_{B}+\alpha_{3}E^{(3)}{}^{A}{}_{B}+\cdots+\alpha
_{n+1}E^{(n+1)}{}^{A}{}_{B}\nonumber\\
&  =\frac{\beta_{0}}{2}T^{(0)}{}^{A}{}_{B}+\frac{\beta_{1}}{8}T^{(1)}{}^{A}%
{}_{B}+\frac{\beta_{2}}{32}T^{(2)}{}^{A}{}_{B}+\cdots+\frac{\beta_{n}%
}{2^{2n+1}}T^{(n)}{}^{A}{}_{B}\ . \label{resultados-principales:ec-lovelock-k}%
\end{align}
Then, introducing (\ref{resultados-principales:ansatz-metrica}) and
(\ref{resultados-principales:ansatz-campo-escalar}) into equation
(\ref{resultados-principales:ec-lovelock-k}), we obtain the equations with
free indices on the string $\mathcal{E}_{\mu\nu}=0$ which imply%

\begin{align}
{\small -}  &  \frac{\tilde{\alpha}_{0}}{2}{\small \delta}_{\nu}^{\mu
}{\small -}\frac{\tilde{\alpha}_{1}}{4}{\small \delta}_{\nu\beta_{1}\beta_{2}%
}^{\mu\alpha_{1}\alpha_{2}}{\small \tilde{R}}^{\beta_{1}\beta_{2}}{\small {}%
{}}_{\alpha_{1}\alpha_{2}}{\small -}\frac{\tilde{\alpha}_{2}}{8}%
{\small \delta}_{\nu\beta_{1}\cdots\beta_{4}}^{\mu\alpha_{1}\cdots\alpha_{4}%
}{\small \tilde{R}}^{\beta_{1}\beta_{2}}{\small {}{}}_{\alpha_{1}\alpha_{2}%
}{\small \tilde{R}}^{\beta_{3}\beta_{4}}{\small {}{}}_{\alpha_{3}\alpha_{4}%
}{\small -\cdots}\nonumber\\
&  {\small -}\frac{\tilde{\alpha}_{n}}{2^{n+1}}{\small \delta}_{\nu\beta
_{1}\cdots\beta_{2n}}^{\mu\alpha_{1}\cdots\alpha_{2n}}{\small \tilde{R}%
}^{\beta_{1}\beta_{2}}{\small {}{}}_{\alpha_{1}\alpha_{2}}{\small \cdots
\tilde{R}}^{\beta_{2n-1}\beta_{2n}}{\small {}{}}_{\alpha_{2n-1}\alpha_{2n}%
}{\small -}\frac{\tilde{\alpha}_{n+1}}{2^{n+2}}{\small \delta}_{\nu\beta
_{1}\cdots\beta_{2n+2}}^{\mu\alpha_{1}\cdots\alpha_{2n+2}}{\small \tilde{R}%
}^{\beta_{1}\beta_{2}}{\small {}{}}_{\alpha_{1}\alpha_{2}}{\small \cdots
\tilde{R}}^{\beta_{2n+1}\beta_{2n+2}}{\small {}{}}_{\alpha_{2n+1}\alpha
_{2n+2}}{\small =0\ ,} \label{resultados-principales:ansatz-ec-mu-nu-l}%
\end{align}
and the equation along the $z$ direction $\mathcal{E}_{zz}$ reads%

\begin{align}
{\small -}  &  \frac{\hat{\alpha}_{0}}{2}{\small -}\frac{\hat{\alpha}_{1}}%
{4}{\small \delta}_{\beta_{1}\beta_{2}}^{\alpha_{1}\alpha_{2}}{\small \tilde
{R}}^{\beta_{1}\beta_{2}}{\small {}{}}_{\alpha_{1}\alpha_{2}}{\small -}%
\frac{\hat{\alpha}_{2}}{8}{\small \delta}_{\beta_{1}\cdots\beta_{4}}%
^{\alpha_{1}\cdots\alpha_{4}}{\small \tilde{R}}^{\beta_{1}\beta_{2}}%
{\small {}{}}_{\alpha_{1}\alpha_{2}}{\small \tilde{R}}^{\beta_{3}\beta_{4}%
}{\small {}{}}_{\alpha_{3}\alpha_{4}}{\small -\cdots}\nonumber\\
&  {\small -}\frac{\hat{\alpha}_{l}}{2^{l+1}}{\small \delta}_{\beta_{1}%
\cdots\beta_{2n}}^{\alpha_{1}\cdots\alpha_{2n}}{\small \tilde{R}}^{\beta
_{1}\beta_{2}}{\small {}{}}_{\alpha_{1}\alpha_{2}}{\small \cdots\tilde{R}%
}^{\beta_{2n-1}\beta_{2n}}{\small {}{}}_{\alpha_{2n-1}\alpha_{2n}}%
{\small -}\frac{\hat{\alpha}_{n+1}}{2^{n+2}}{\small \delta}_{\beta_{1}%
\cdots\beta_{2n+2}}^{\alpha_{1}\cdots\alpha_{2n+2}}{\small \tilde{R}}%
^{\beta_{1}\beta_{2}}{\small {}{}}_{\alpha_{1}\alpha_{2}}{\small \cdots
\tilde{R}}^{\beta_{2n+1}\beta_{2n+2}}{\small {}{}}_{\alpha_{2n+1}\alpha
_{2n+2}}{\small =0\ .} \label{resultados-principales:ansatz-ec-z-z-l}%
\end{align}
Here we have introduced for simplicity both sets of shifted gravitational
couplings $\tilde{\alpha}_{k}$ and $\hat{\alpha}_{k}$, given respectively by%
\begin{equation}
\tilde{\alpha}_{k}=\left\{
\begin{array}
[c]{cc}%
\alpha_{k}-\frac{\beta_{k}}{2^{k+1}}c_{k}^{2} & \text{for }0\leq k\leq n\\
\alpha_{k} & \text{for }k=n+1
\end{array}
\right.  \ ,
\end{equation}
and%
\begin{equation}
\hat{\alpha}_{k}=\left\{
\begin{array}
[c]{cc}%
\alpha_{k}+\frac{\beta_{k}}{2^{k+1}}c_{k}^{2} & \text{for }0\leq k\leq n\\
\alpha_{k} & \text{for }k=n+1
\end{array}
\right.  \ .
\end{equation}
In general, the compatibility of the equations $\tilde{g}^{\mu\nu}%
\mathcal{E}_{\mu\nu}=0$ and $\mathcal{E}_{zz}=0$ will induce further
restrictions on the metric function $f\left(  r\right)  $ than those that come
only from the equations on the brane $\mathcal{E}_{\mu\nu}=0$. We can
circumvent this clash by imposing the off-shell requirement $\tilde{g}^{\mu
\nu}\mathcal{E}_{\mu\nu}\sim\mathcal{E}_{zz}$ and introducing a
proportionality constant $\xi$ one obtains%

\begin{equation}
\hat{\alpha}_{k}=\xi\left(  d-2k\right)  \tilde{\alpha}_{k}\text{ for }0\leq
k\leq n\ , \label{resultados-principales:traza-alpha-l-l}%
\end{equation}
and%
\begin{equation}
\xi\left(  d-2\left(  n+1\right)  \right)  =1\ .
\end{equation}
These equations allow to fix the proportionality constant $\xi$ as well as the
$n$ integration constants $c_{k}$ for all the scalars in the following manner%
\begin{equation}
c_{k}^{2}=\frac{2^{k+1}\left(  n+1-k\right)  }{\left(  d-k-n-1\right)  }%
\frac{\alpha_{k}}{\beta_{k}}\text{ with }0\leq k\leq n\ .
\end{equation}
Then, the equations on the brane, reduce to an effective Wheeler-like
polynomial that reads%

\begin{align}
m & =\left( d-2\right) V^{\left( K\right) }_{d-2}\sum^{n+1}_{k=0}\frac{\left(
d-2n-2\right) }{\left( d-k-n-1\right) }\frac{\left( d-3\right) !}{\left(
d-2k-1\right) !}\alpha_{k}r^{d-2k-1}\left( K-f\left( r\right) \right)
^{k}.\label{ec-termo-general:polinomio}%
\end{align}
where $m$ is an integration constant and $V^{\left( K\right) }_{d-2}$ stands
for the volume of manifold with line element $d\Sigma_{K,d-2}$. This leads to
a black string in a general Lovelock theory of gravity that can be
asymptotically flat or asymptotically AdS depending of whether we include a
cosmological term in the action. The temperature and the entropy of the
corresponding black hole are respectively given by%

\begin{align}
T & =\frac{1}{4\pi}\frac{\sum^{n+1}_{k=0}\frac{\left( d-2k-1\right) \left(
d-2n-2\right) }{\left( d-k-n-1\right) }\frac{\left( d-3\right) !}{\left(
d-2k-1\right) !}\alpha_{k}r^{d-2k-2}_{+}K^{k}}{\sum^{n+1}_{p=0}\frac{p\left(
d-2n-2\right) }{\left( d-p-n-1\right) }\frac{\left( d-3\right) !}{\left(
d-2p-1\right) !}\alpha_{p}r^{d-2p-1}_{+}K^{p-1}}%
.\label{ec-termo-general:temperatura}%
\end{align}

and%

\begin{align}
s & =4\pi\sum^{n+1}_{k=0}\frac{k\left( d-2n-2\right) }{\left(
d-k-n-1\right) }\frac{\left( d-2\right) }{\left( d-2k\right) }\frac{\left(
d-3\right) !}{\left( d-2k-1\right) !}\alpha_{k}r^{d-2k}_{+}K^{k-1}V^{\left(
K\right) }_{d-2}.\label{ec-termo-general:entropia}%
\end{align}

and it can be checked that the first law $dm=Tds$ is fulfilled, which lead to
the interpretation of the integration constant $m$ as the mass density.

\bigskip

The compatibility of the equations $\tilde{g}^{\mu\nu}\mathcal{E}_{\mu\nu}=0$
and $\mathcal{E}_{zz}=0$ is ensured for a general metric on the transverse
section of the string, and the scalars%
\begin{equation}
\phi_{\left(  k\right)  }\left(  z\right)  =\left(  \frac{2^{k+1}\left(
n+1-k\right)  }{\left(  d-k-n-1\right)  }\frac{\alpha_{k}}{\beta_{k}}\right)
^{1/2}z\ ,
\end{equation}
allow to cylindrically extend any solution of Lovelock theory from dimension
$d$ to dimension $d+1$. This will be exploited in the next section.

\section{Extra physically interesting solutions}

As usual in Lovelock theories, from the polynomial equation
(\ref{ec-termo-general:polinomio}) when the integration constant $m$ vanishes,
one will obtain $n$ different solutions for the metric function $f\left(
r\right)  $, leading to $n$ different constant curvature solutions on the
transverse section of the string. When the curvature radii of these solutions
coincide, it is natural to expect an enlargement of the space of solutions of
the theory, as it occurs in vacuum. To see this in a concrete example, let us
revisit the Einstein-Gauss-Bonnet theory in six dimensions%
\begin{equation}
I=\int d^{6}x\sqrt{-g}\left(  R-2\Lambda_{0}+\alpha\mathcal{L}_{GB}-\frac
{1}{2}g_{AB}\partial^{A}\psi\partial^{B}\psi+\frac{\gamma}{2}G_{AB}%
\partial^{A}\chi\partial^{B}\chi\right)  \ .
\end{equation}
This theory admits a cylindrically extended wormhole solution with line
element given by%
\begin{equation}
ds^{2}=l^{2}\left[  -\cosh^{2}\left(  \rho-\rho_{0}\right)  dt^{2}+d\rho
^{2}+\cosh^{2}\rho d\Sigma_{3}^{2}\right]  +dz^{2}\ ,\label{wormstring}%
\end{equation}
dressed by the scalars%
\begin{equation}
\psi\left(  z\right)  =\frac{2\sqrt{3}}{l}z\text{ and }\chi\left(  z\right)
=\sqrt{\frac{2}{\gamma}}z\ ,\label{scalars}%
\end{equation}
provided%
\begin{equation}
l^{2}=8\alpha\text{\ ,}\label{lsq1}%
\end{equation}
and%
\begin{equation}
\Lambda_{0}\alpha=-\frac{9}{16}\ .\label{noncs}%
\end{equation}
The metric (\ref{wormstring}) describes the cylindrical extension of a
wormhole metric, where $\rho_{0}$ is an arbitrary integration constant and
$d\Sigma_{3}$ is the line element of an Euclidean, three dimensional space
with Ricci scalar equals to $-6$ \cite{DOT}. In the original five-dimensional
formulation, the wormhole solution \cite{DOT} exist for Einstein-Gauss-Bonnet
theory in the Chern-Simons case in which the five dimensional theory, when
formulated in first order formalism, have an enlargement of the local symmetry
group from the usual Lorentz group to an AdS local group, and the vielbein and
spin connections belong to an AdS gauge connection \cite{HZ}. From the point
of view of a dimensional reduction, the scalars considered here, allow to
start with a generic six-dimensional theory and obtained an effective
Chern-Simons theory in dimension five.

\bigskip

It is interesting to notice that even though the latter solution has a simple
warped structure in five dimensions, one can also embed in a similar manner, a
rotating solution of the effective five dimensional Chern-Simons theory, into
dimension six. Explicitly the following metric is a solution of the field
equations:%
\begin{equation}
ds^{2}=d\bar{s}_{L}^{2}+\left(  \frac{1}{L^{2}}-\frac{1}{l^{2}}\right)
\rho^{2}\left(  r,\mu\right)  k_{\alpha}k_{\beta}dx^{\alpha}dx^{\beta}+dz^{2},
\label{KSstring}%
\end{equation}
where $d\bar{s}_{L}$ is the line element of AdS$_{5}$ with curvature radius
$L$, in double oblate coordinates, i.e.%
\begin{align*}
d\bar{s}_{L}^{2}  &  =-\left(  1+\frac{r^{2}}{L^{2}}\right)  \frac
{\Delta\left(  \mu\right)  }{\Xi_{a}\Xi_{b}}dt^{2}+\frac{r^{2}\rho^{2}dr^{2}%
}{\left(  1+\frac{r^{2}}{L^{2}}\right)  \left(  r^{2}+a^{2}\right)  \left(
r^{2}+b^{2}\right)  }+\frac{\rho^{2}d\mu^{2}}{\Delta\left(  \mu\right)
\left(  1-\mu^{2}\right)  }\\
&  +\frac{(r^{2}+a^{2})\left(  1-\mu^{2}\right)  }{\Xi_{a}}d\phi^{2}%
+\frac{(r^{2}+b^{2})\mu^{2}}{\Xi_{b}}d\psi^{2}\ ,
\end{align*}
and $k=k_{\alpha}dx^{\alpha}$ reads%
\[
k=\frac{\Delta(\mu)}{\Xi_{a}\Xi_{b}}dt+\frac{r^{2}\rho^{2}dr}{\left(
1+\frac{r^{2}}{L^{2}}\right)  \left(  r^{2}+a^{2}\right)  \left(  r^{2}%
+b^{2}\right)  }+\frac{a\left(  1-\mu^{2}\right)  }{\Xi_{a}}d\phi+\frac
{b\mu^{2}}{\Xi_{b}}d\psi\ ,
\]
where $\Xi_{a}=1-\frac{a^{2}}{L^{2}},\ \Xi_{b}=1-\frac{b^{2}}{L^{2}}%
,\ \Delta\left(  \mu\right)  =\Xi_{a}\mu^{2}+\Xi_{b}\left(  1-\mu^{2}\right)
$, and $\rho^{2}(r,\mu)=r^{2}+a^{2}\mu^{2}+b^{2}\left(  1-\mu^{2}\right)  \ $.
The scalar fields are given by (\ref{scalars}) and the relations (\ref{lsq1})
and (\ref{noncs}), hold. Note that $l$ is a constant defined by the theory,
while $L$ is an \ integration constant as well as $a$ and $b$. The latter are
the so-called oblateness parameters. Since the vector field $k$ defines a null
and geodesic congruence of the background metric $d\bar{s}_{L}$, the metric
(\ref{KSstring}) is a cylindrical oxidation of a Kerr-Schild metric in
AdS$_{5}$. The five dimensional metric in the section of the string defines
the first known exact, analytic, rotating solution of Einstein-Gauss-Bonnet in
five dimensions and it was originally found in \cite{Anabalon:2009kq}. It was
also latter proven to be of the non-circular type in \cite{Anabalon:2010ns}
\footnote{For a recent extension of the rotating metric in Chern-Simons
theories in odd dimensions, within the Lovelock family in vacuum, see
\cite{cvetic}.}.

\section{Final remarks}
Until now in higher curvature gravity, exact and homogeneous black string solutions have been constructed only for special values of the coupling constants \cite{GOT,Kastor:2006vw} also including $p$-form fields \cite{pform}, and the general problem of the construction of Lovelock branes was studied in \cite{Kastor:2017knv}, and for arbitrary values of the couplings only numerical or perturbative solutions were available (see e.g. \cite{pertgb}). In this paper we have constructed new, exact, homogeneous black strings in
arbitrary Lovelock theories. The solutions are supported by scalar fields with
non-minimal kinetic couplings constructed with Lovelock tensors, ensuring that
the field equations are of second order. These scalars have been previously
considered as part of the higher dimensional extension of Horndeski theories
in \cite{Charmousis:2015txa,GAO}. Here, the scalars being dependent only on
the coordinate along the extended direction, turn out to be linear, and the
proportionality constant gets fixed by the requiring the compatibility of the
whole system. The transverse section of the string can be asymptotically flat
or $AdS$. For concreteness we computed the thermodynamic quantities and showed
that the entropy of the black string receives a contribution from the matter
part. This is interesting because the pattern of transitions between black
string and black hole can change due to the presence of the new Horndeski fields. The gravitational stability of the black string and $p$-branes in the presence of a single quadratic or cubic Lovelock term has been studied in \cite{sta1}-\cite{sta3}, while the effect of quartic corrections coming from M-theory have been explored in \cite{hyakutake}. It is interesting to mention that the Large D approach \cite{est} allows to keep all the terms in the Einstein-Gauss-Bonnet Lagrangian \cite{largedegb}. 

For simplicity we focus on the case with a single extended direction, but this
construction also works with $p$-branes. For example, in the
Einstein-Gauss-Bonnet theory, with a cosmological constant, one would have to
consider $p$ minimally coupled scalars mimicking such of reference
\cite{Cisterna:2017qrb}, as well as $p$ scalars with Einstein-kinetic
couplings which would also turn out to be linear and depending on a single
extended direction. From this, the extension to arbitrary Lovelock theories,
with flat $p$-branes is clear.

Interpreting our solutions as compactifications with non-trivial scalar fluxes
along the extended direction, one obtains an effective Lovelock theory induced
on the brane. We exploited this idea to construct cylindrically extended
solution with wormholes on the transverse section, which are asymptotically
AdS$_{5}\times R$, in both directions. Within the same realm we also
constructed a cylindrical oxidation of the rotating spacetime of
Einstein-Gauss-Bonnet gravity, constructed from a Kerr-Schild ansatz in
\cite{Anabalon:2009kq}. The wormhole solution can of course be extended to
AdS$_{2n-1}\times R$ provided one considers Lovelock theory with all the
possible terms in dimension $2n$. In such case, the wormhole on the string
will be the one reported in \cite{DOT}.

\bigskip

\section{Acknowledgments}

The authors would like to acknowledge E. Babichev, C. Charmousis, N. Grandi
and J. Rocha for valuable comments and remarks. J.O. is partially supported by
FONDECYT grant 1181047. A. C. is supported by Fondo Nacional de Desarrollo
Cient\'{\i}fico y Tecnol\'{o}gico Grant No. 11170274 and Proyecto Interno Ucen
I+D-2016, CIP2016.

\end{document}